\documentclass[aps,twocolumn,superscriptaddress,showpacs]{revtex4}%
\usepackage{amsfonts}
\usepackage{amsmath}
\usepackage{amssymb}
\usepackage{graphicx}%
\providecommand{\U}[1]{\protect\rule{.1in}{.1in}}
\providecommand{\U}[1]{\protect\rule{.1in}{.1in}}

\begin{document}
\title{Perfect optical nonreciprocity in a double-cavity optomechanical system}
\author{\ Xiao-Bo Yan}
\email{xiaoboyan@126.com}
\affiliation{College of Electronic Science, Northeast Petroleum University, Daqing 163318, China}
\author{\ He-Lin Lu}
\affiliation{Department of Physics, Yunnan Minzu University, Kunming 650500, China}
\author{\ Feng Gao}
\affiliation{College of Science, Shenyang Aerospace University, Shenyang 110136, China}
\author{\ Liu Yang}
\email{lyang@hrbeu.edu.cn}
\affiliation{College of Automation, Harbin Engineering University, Harbin 150001, China}

\date{\today }



\begin{abstract}
Nonreciprocal devices are indispensable for building quantum networks and ubiquitous in modern communication
technology. Here, we propose to take advantage of the interference between optomechanical
interaction and linearly-coupled interaction to realize optical nonreciprocal transmission in
a double-cavity optomechanical system. Particularly, we have derived essential conditions for perfect
optical nonreciprocity and analysed properties of the optical nonreciprocal transmission. These results can be used to control optical transmission in quantum information processing.

\end{abstract}
\maketitle

\section{Introduction}

Nonreciprocal optical devices, such as isolators and circulators,
allow transmission of signals to exhibit different characteristics if source and observer are interchanged. They are essential to several applications in quantum signal
processing and communication, as they can suppress spurious modes and
unwanted signals \cite{Jalas2013}. For example, they can protect devices from noise emanating from readout
electronics in quantum superconducting circuits. To
violate reciprocity and obtain asymmetric transmission, breaking time-reversal symmetry is required in any such device.
Traditionally, nonreciprocal transmission has
relied on applied magnetic bias fields to break time-reversal
symmetry and Lorentz reciprocity \cite{Hogan,Aplet}. These conventional devices are typically bulky and incompatible with ultra-low loss superconducting circuits.
Many alternative methods recently have been proposed to replace conventional nonreciprocal schemes, such as usage of coupled-mode systems \cite{Ranzani2015,BingHe2018}, Brillouin scattering \cite{Dong2015}, and spatiotemporal
modulation of the refractive index \cite{Fang2012}. These schemes are particularly promising because
they can be integrated on-chip with existing superconducting
technology.

In recent years, the rapidly growing field
of cavity optomechanics \cite{Aspelmeyer2014,Hu2013,Zhang2011}, where optical fields and mechanical resonators are coupled
through radiation pressure, has shown
promising potential for applications in quantum information
processing and communication.
So far, many interesting quantum phenomena have been studied in this field, such as mechanical ground-state cooling \cite{Marquardt2007,Wilson-Rae2007,BingHe2017},
optomechanically induced transparency \cite{Agarwal2010,Weis2010,CDong2013,YCLiu2017,HXiong2018,HZhang2018,Yan2015}, entanglement \cite{sun2017,Tian2013,Deng2015,Deng2016,Yan2017,Yan2018,Zhong2018,Liu2019}, nonlinear effects \cite{Borkje2013,LuXY2013}, and coherent perfect absorption \cite{Yan2014,Agarwal2014,Zhao2019}. Very recently, it has been realized that optomechanical coupling can lead to nonreciprocal transmission and
optical isolation. In Refs. \cite{Manipatruni2009,Rabl2012,ZqWang2015,Miri2017,Sounas2017}, nonreciprocal optical responses are theoretically predicted through optomechanical interactions, and nonreciprocal transmission spectra were recently observed in Refs. \cite{Ruesink2016,Shen2016,Peterson2017,Bernier2017,Barzanjeh2017,Fang2017,Ruesink2018}. In Refs. \cite{Tian2017,Xu2015,Xu2016,Xia2019}, it
was recognized that the mechanically-mediated quantum-state transfer
between two cavity modes can be made nonreciprocal with
suitable optical driving.
In Ref. \cite{Malz2018}, nonreciprocal quantum-limited amplification
of microwave signals has been proposed in an optomechanical system. Besides, in Refs. \cite{Habraken2012,Seif2018}, phonon circulators and thermal diodes are theoretically predicted through optomechanical coupling.
In most of these references, perfect optical nonreciprocity can be achieved under the conditions of equal damping rate (mechanical damping rate $\gamma$ is equal to cavity damping rate $\kappa$) or nonreciprocal phase difference $\theta=\pm\frac{\pi}{2}$.

\begin{figure}[b]
\centering\includegraphics[width=0.4\textwidth]{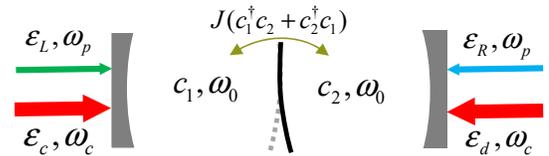}
\caption{A
double-cavity optomechanical system with a mechanical resonator interacted
with two cavities. Two strong coupling fields (probe fields) with amplitudes $\varepsilon_{c}$ and
$\varepsilon_{d}$ ($\varepsilon_{L}$ and $\varepsilon_{R}$) are used to drive
the system from the left and right fixed mirror respectively. Meanwhile, the two cavities are linearly coupled to each other with coupling strength $J$.}
\label{f1}
\end{figure}

Here, we show that perfect optical nonreciprocity can be achieved under more general conditions, using the example of a double-cavity system in Fig. 1. This setup has been realized in several
recent experiments \cite{Thompson2008,Jayich2008,Sankey2010}, and quantum nonlinearity \cite{Ludwig2012} has been theoretically studied in this setup.
With this simple model, we can easily capture the essential mechanisms about nonreciprocity, i.e., quantum interference of signal transmission between two possible paths corresponding to two interactions (optomechanical interaction and linearly-coupled interaction). From the expressions of output fields, we derive essential conditions
to achieve perfect optical nonreciprocity, and find some interesting results. One of them is that mechanical decay rate does not influence the appearance of perfect optical nonreciprocity, which means perfect optical nonreciprocity can still occur in the realistic parameter regime ($\gamma\ll\kappa$) in cavity optomechanics. Another interesting result is that perfect optical nonreciprocity can be achieved with any phase difference $\theta$ ($\theta\neq0,\pi$) as long as rotating wave approximation is valid. We believe the results of this paper can be used to control optical transmission in modern communication
technology.

\section{System model and equations}

The system we considered here is depicted in Fig. 1, where a mechanical membrane is placed in the middle of an
optical cavity. The operators $c_{1}$ and $c_{2}$ denote geometrically distinct optical modes with same frequency $\omega_{\text{0}}$ and decay rates $\kappa_{1}$ and
$\kappa_{2}$ respectively. The coupling $J$ describes photon tunneling through the membrane, and the interaction between the two cavities is described by $\hbar J(c^{\dag}_{1}c_{2}+c^{\dag}_{2}c_{1})$. The mechanical
resonator with frequency $\omega_{\text{m}}$ and decay rate
$\gamma$ is described by operator $b$. Two strong coupling fields (probe fields) with same frequency
$\omega_{c}$ ($\omega_{p}$) and amplitudes $\varepsilon_{c}$ and
$\varepsilon_{d}$ ($\varepsilon_{L}$ and $\varepsilon_{R}$) are used to drive
the double-cavity system from the left and right fixed mirror respectively.
Then the total Hamiltonian in the
rotating-wave frame of coupling frequency $\omega_{c}$ can be written as
($\hbar=1$)
\begin{align}
H  &  =\Delta_{c}(c_{1}^{\dag}c_{1}+c_{2}^{\dag}c_{2})+\omega_{\text{m}}
b^{\dag}b+g_{0}(c_{2}^{\dag}c_{2}-c_{1}^{\dag}c_{1})(b^{\dag
}+b)\nonumber\\
&  +J(c_{1}^{\dag}c_{2}+c_{2}^{\dag}c_{1})+i(\varepsilon_{c}c_{1}^{\dag
}-\varepsilon^{\ast}_{c}c_{1})+i(\varepsilon_{d}c_{2}^{\dag}-\varepsilon
^{\ast}_{d}c_{2})\nonumber\\
&  +i\varepsilon_{L}(c_{1}^{\dag}e^{-i\delta t}-c_{1}e^{i\delta t}
)+i\varepsilon_{R}(c_{2}^{\dag}e^{-i\delta t}-c_{2}e^{i\delta t}) .
\label{Eq1}
\end{align}
Here, $\Delta_{c}=\omega_{0}-\omega_{c}$ ($\delta=\omega_{p}-\omega_{c}$) is the detuning between cavity modes (probe fields)
and coupling fields, and $g_{0}$ is the single photon coupling
constant between mechanical resonator and optical modes.

The dynamics of the system is described by the quantum Langevin equations for
the relevant operators of the mechanical and optical modes
\begin{align}
\dot{c}_{1}  &  =-[i\Delta_{c}+\frac{\kappa_{1}}{2}-ig_{0}(b^{\dag}%
+b)]c_{1}+\varepsilon_{c}+\varepsilon_{L}e^{-i\delta t}-iJc_{2},\nonumber\\
\dot{c}_{2}  &  =-[i\Delta_{c}+\frac{\kappa_{2}}{2}+ig_{0}(b^{\dag}%
+b)]c_{2}+\varepsilon_{d}+\varepsilon_{R}e^{-i\delta t}-iJc_{1},\nonumber\\
\dot{b}  &  =-i\omega_{m}b-\frac{\gamma}{2}b-ig_{0}(c_{2}^{\dag}%
c_{2}-c_{1}^{\dag}c_{1}). \label{Eq2}%
\end{align}
In the absence of probe fields $\varepsilon_{L}$, $\varepsilon_{R}$ and with
the factorization assumption $\langle bc_{i}\rangle=\langle b\rangle\langle
c_{i}\rangle$, we can obtain the steady-state mean values
\begin{align}
\langle b\rangle &  =b_{s}=\frac{-ig_{0}(\left\vert c_{2s}\right\vert
	^{2}-\left\vert c_{1s}\right\vert ^{2})}{\frac{\gamma}{2}+i\omega_{m}%
},\nonumber\\
\left\langle c_{1}\right\rangle  &  =c_{1s}=\frac{(\frac{\kappa_{2}}%
	{2}+i\Delta_{2})\varepsilon_{c}-iJ\varepsilon_{d}}{J^{2}+(\frac{\kappa_{1}}%
	{2}+i\Delta_{1})(\frac{\kappa_{2}}{2}+i\Delta_{2})},\nonumber\\
\left\langle c_{2}\right\rangle  &  =c_{2s}=\frac{(\frac{\kappa_{1}}{2}%
	+i\Delta_{1})\varepsilon_{d}-iJ\varepsilon_{c}}{J^{2}+(\frac{\kappa_{1}}%
	{2}+i\Delta_{1})(\frac{\kappa_{2}}{2}+i\Delta_{2})} \label{Eq3}%
\end{align}
with $\Delta_{1,2}=\Delta_{c}\mp g_{0}(b_{s}+b_{s}^{\ast})$ denoting the
effective detunings between cavity modes and coupling fields. In the presence
of both probe fields, however, we can write each operator as the sum of its
mean value and its small fluctuation, i.e., $b=b_{s}+\delta b$, $c_{1}%
=c_{1s}+\delta c_{1}$, $c_{2}=c_{2s}+\delta c_{2}$ to solve Eq. (2) when both
coupling fields are sufficiently strong. Then keeping only the linear terms of
fluctuation operators and moving into an interaction picture by introducing
$\delta b\rightarrow\delta be^{-i\omega_{m}t}$, $\delta c_{1}\rightarrow\delta
c_{1}e^{-i\Delta_{1}t}$, $\delta c_{2}\rightarrow\delta c_{2}e^{-i\Delta_{2}%
	t}$, we obtain the linearized quantum Langevin equations
\begin{align}
\delta\dot{c}_{1}  &  =-\frac{\kappa_{1}}{2}\delta c_{1}+iG_{1}(\delta
b^{\dag}e^{i(\omega_{m}+\Delta_{1})t}+\delta be^{-i(\omega_{m}-\Delta_{1}%
	)t})\nonumber\\
&  +\varepsilon_{L}e^{-i(\delta-\Delta_{1})t}-iJ\delta c_{2}e^{i(\Delta_{1}-\Delta_{2})t},\nonumber\\
\delta\dot{c}_{2}  &  =-\frac{\kappa_{2}}{2}\delta c_{2}-iG_{2}e^{i\theta
}(\delta b^{\dag}e^{i(\omega_{m}+\Delta_{2})t}+\delta be^{-i(\omega_{m}%
	-\Delta_{2})t})\nonumber\\
&  +\varepsilon_{R}e^{-i(\delta-\Delta_{2})t}-iJ\delta c_{1}e^{i(\Delta_{2}-\Delta_{1})t},\nonumber\\
\delta\dot{b}  &  =-\frac{\gamma}{2}\delta b+iG_{1}(\delta c_{1}%
e^{i(\omega_{m}-\Delta_{1})t}+\delta c^{\dag}_{1}e^{i(\omega_{m}+\Delta_{1}%
	)t})\nonumber\\
&  -iG_{2}(e^{-i\theta}\delta c_{2}e^{i(\omega_{m}-\Delta_{2})t}+e^{i\theta
}\delta c^{\dag}_{2}e^{i(\Delta_{2}+\omega_{m})t}) \label{Eq4}%
\end{align}
with $G_{1}=g_{0}c_{1s}$ and $G_{2}=g_{0}c_{2s}e^{-i\theta}$. The phase difference $\theta$ between effective optomechanical coupling $g_{0}c_{1s}$ and
$g_{0}c_{2s}$ can be controlled by adjusting the coupling fields amplitudes $\varepsilon_{c}$ and $\varepsilon_{d}$ according to Eq. (3). It will be seen that the phase difference $\theta$ is a critical factor to attain optical nonreciprocity. Without
loss of generality, we take $G_{i}$ and $J$ as \emph{positive} number (not \emph{negative} to avoid introducing unimportant phase difference $\pi$).

If each coupling field drives one cavity mode at the mechanical red sideband
$(\Delta_{1}\approx\Delta_{2}\approx\omega_{m})$, and the mechanical frequency $\omega_{m}$ is much larger than
$g_{0}|c_{1s}|$ and $g_{0}|c_{2s}|$, then Eq. (4) will be simplified to
\begin{align}
\delta\dot{c}_{1}  &  =-\frac{\kappa_{1}}{2}\delta c_{1}+iG_{1}\delta
b-iJ\delta c_{2}+\varepsilon_{L}e^{-ixt},\nonumber\\
\delta\dot{c}_{2}  &  =-\frac{\kappa_{2}}{2}\delta c_{2}-iG_{2}e^{i\theta
}\delta b-iJ\delta c_{1}+\varepsilon_{R}e^{-ixt},\nonumber\\
\delta\dot{b}  &  =-\frac{\gamma}{2}\delta b+iG_{1}\delta c_{1}%
-iG_{2}e^{-i\theta}\delta c_{2} \label{Eq5}%
\end{align}
with $x=\delta-\omega_{m}$. For simplicity, we set equal damping rate $\kappa_{1}=\kappa_{2}=\kappa$ and equal coupling $G_{1}=G_{2}=G$ in the following (actually, it can be proven that $G_{1}$ must equal $G_{2}$ if $\kappa_{1}=\kappa_{2}$ when the system exhibits perfect optical nonreciprocity).

By assuming $\delta
s=\delta s_{+}e^{-ixt}+\delta s_{-}e^{ixt}$ ($s=b,c_{1},c_{2}$), we can solve Eq. (5) as follows
\begin{align}
\delta b_{+}  &  =\frac{4G[(i\kappa_{x}-2Je^{-i\theta})\varepsilon
	_{L}+(2J-i\kappa_{x}e^{-i\theta})\varepsilon_{R}]}{8G^{2}\kappa_{x}+(4J^{2}+\kappa^{2}
	_{x})\gamma_{x}+16iG^{2}J\cos\theta},\nonumber\\
\delta c_{1+}  &  =\frac{2(4G^{2}+\gamma_{x}\kappa_{x})\varepsilon_{L}%
	+(8G^{2}e^{-i\theta}-4iJ\gamma_{x})\varepsilon_{R}}{8G^{2}\kappa_{x}+(4J^{2}+\kappa^{2}
	_{x})\gamma_{x}+16iG^{2}J\cos\theta},\nonumber\\
\delta c_{2+}  &  =\frac{2(4G^{2}+\gamma_{x}\kappa_{x})\varepsilon_{R}%
	+(8G^{2}e^{i\theta}-4iJ\gamma_{x})\varepsilon_{L}}{8G^{2}\kappa_{x}+(4J^{2}+\kappa^{2}
	_{x})\gamma_{x}+16iG^{2}J\cos\theta} \label{Eq6}%
\end{align}
where $\gamma_{x}=\gamma-2ix$, $\kappa_{x}=\kappa-2ix$, and $\delta s_{-}=0$.

To study optical
nonreciprocity, we must study the output optical fields $\varepsilon^{out}%
_{L}$ and $\varepsilon^{out}_{R}$ which can be obtained according to the
input-output relation \cite{Yan2014,Agarwal2014,Walls}
\begin{align}
\varepsilon^{out}_{L}+\varepsilon^{in}_{L}e^{-ixt}  &  =\sqrt{\kappa}\delta
c_{1}\nonumber\\
\varepsilon^{out}_{R}+\varepsilon^{in}_{R}e^{-ixt}  &  =\sqrt{\kappa}\delta c_{2},
\label{Eq7}%
\end{align}
here, $\varepsilon^{in}_{L,R}=\varepsilon_{L,R}/\sqrt{\kappa}$. Still following the assumption $\delta s=\delta s_{+}e^{-ixt}+\delta s_{-}e^{ixt}$, the output fields can be obtained as
\begin{align}
\varepsilon^{out}_{L+}  &  =\sqrt{\kappa}\delta c_{1+}-\varepsilon_{L}/\sqrt{\kappa}\nonumber\\
\varepsilon^{out}_{R+}  &  =\sqrt{\kappa}\delta c_{2+}-\varepsilon_{R}/\sqrt{\kappa}
\label{Eq8}%
\end{align}
and $\varepsilon^{out}_{L-}=\varepsilon^{out}_{R-}=0$.

\section{Perfect optical nonreciprocity}

Perfect optical nonreciprocity can be achieved if transmission amplitudes $T_{i\rightarrow j}$ ($i,j=L,R$) satisfy
\begin{align}
T_{L\rightarrow R}=\left\vert \frac{\varepsilon_{R}^{out}}{\varepsilon^{in}_{L}}\right\vert
_{\varepsilon^{in}_{R}=0}=1,T_{R\rightarrow L}=\left\vert \frac{\varepsilon_{L}^{out}}{\varepsilon^{in}
	_{R}}\right\vert _{\varepsilon^{in}_{L}=0}=0,\tag{9a}
\end{align}
or
\begin{align}
T_{L\rightarrow R}=\left\vert \frac{\varepsilon_{R}^{out}}{\varepsilon^{in}_{L}}\right\vert
_{\varepsilon^{in}_{R}=0}=0,T_{R\rightarrow L}=\left\vert \frac{\varepsilon_{L}^{out}}{\varepsilon^{in}
	_{R}}\right\vert _{\varepsilon^{in}_{L}=0}=1.\tag{9b}
\end{align}
It means that the input signal from one side can be completely transmitted to
the other side, but not vice versa. What the Eq. (9a) and (9b) represent is the two different directions of isolation. Here, we just discuss the case of Eq. (9a), as the case of Eq. (9b) is similar. The subscript $\varepsilon^{in}_{R/L}=0$ indicates there is not signal injected into the system from right/left side. We omit the subscripts because, in general, nonreciprocity is only related to one-way input, and write transmission amplitudes $T_{i\rightarrow j}$ as $T_{ij}$ for simplicity in the following.

According to Eq. (6) and Eq. (8), the two optical output fields can be
obtained as
\begin{align}
\frac{\varepsilon_{R+}^{out}}{\varepsilon^{in}_{L}}&
=\frac{4\kappa(2G^{2}e^{i\theta}-iJ\gamma_{x})}{8G^{2}\kappa_{x}+(4J^{2}+\kappa^{2}
	_{x})\gamma_{x}+16iG^{2}J\cos\theta},\nonumber\\
\frac{\varepsilon_{L+}^{out}}{\varepsilon^{in}_{R}}&
=\frac{4\kappa(2G^{2}e^{-i\theta}-iJ\gamma_{x})}{8G^{2}\kappa_{x}+(4J^{2}+\kappa^{2}
	_{x})\gamma_{x}+16iG^{2}J\cos\theta}.\tag{10}\label{Eq10b}%
\end{align}
When $\theta=n\pi$ ($n$ is an integer), the two output fields are equal, which indicates the photon transmission is reciprocal. But in the other cases, where $\theta\neq n\pi$, the system will exhibit a nonreciprocal response. It can be clearly seen from the numerator of Eq. (10) that the optical nonreciprocity comes from quantum interference between the optomechanical interaction $G$ and the linearly-coupled interaction $J$.

With Eq. (10), we find perfect optical nonreciprocity Eq. (9a) can be achieved only when
\begin{align}
J=-\frac{e^{\mp i\theta}(\gamma\cot\theta\pm i\kappa)}{2}\tag{11}
\end{align}
which can take positive real number only if
\begin{align}
\theta=-\frac{\pi}{2},\tag{12a}
\end{align}
or
\begin{align}
\kappa=\gamma. \tag{12b}
\end{align}
In the following, we will discuss perfect optical nonreciprocity in two cases, Eq. (12a) and (12b), respectively.

\subsection{Phase difference $\theta=-\frac{\pi}{2}$}

\begin{figure}[b]
\centering\includegraphics[width=0.48\textwidth]{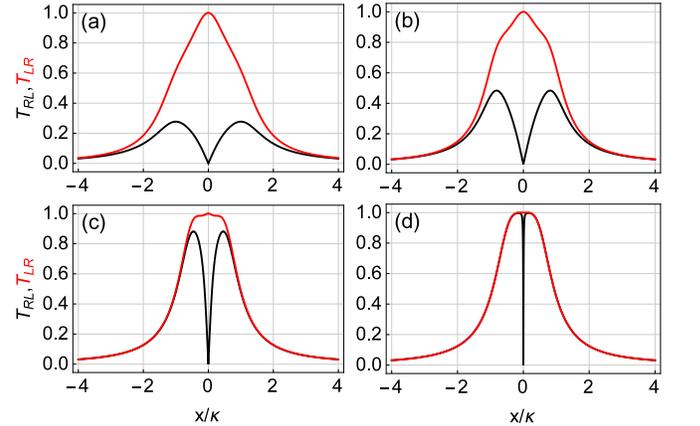} \caption{ Transmission amplitudes $T_{LR}$ (red line) and $T_{RL}$ (black line) are plotted vs normalized detuning $x/\kappa$ for different mechanical decay rate: (a) $\gamma/\kappa$=2, (b) $\gamma/\kappa$=1, (c) $\gamma/\kappa$=1/5, and (d) $\gamma/\kappa$=1/100. The coupling strengths $J=\frac{\kappa}{2}$ and $G=\frac{\sqrt{\kappa\gamma}}{2}$ according to Eq. (14).}%
\label{Fig2}%
\end{figure}

With nonreciprocal phase difference $\theta=-\frac{\pi}{2}$, the two optical output fields in Eq. (10) now become
\begin{align}
\frac{\varepsilon_{R+}^{out}}{\varepsilon^{in}_{L}}&
=\frac{-4i\kappa(2G^{2}+J\gamma_{x})}{8G^{2}\kappa_{x}+(4J^{2}+\kappa^{2}
	_{x})\gamma_{x}},\nonumber\\
\frac{\varepsilon_{L+}^{out}}{\varepsilon^{in}_{R}}&
=\frac{4i\kappa(2G^{2}-J\gamma_{x})}{8G^{2}\kappa_{x}+(4J^{2}+\kappa^{2}
	_{x})\gamma_{x}}.\tag{13}%
\end{align}
According to Eq. (13), the perfect optical nonreciprocity Eq. (9a) can be achieved only when
\begin{align*}
x&=0, \nonumber\\
J&=\frac{\kappa}{2},\nonumber\\
G&=\frac{\sqrt{\kappa\gamma}}{2}.\tag{14}
\end{align*}
It is surprising that there is not any restriction on mechanical decay rate $\gamma$ in Eq. (14). In other words, mechanical decay rate $\gamma$ does not influence perfect optical nonreciprocity, which means that perfect optical nonreciprocity can still occur even in the case of $\gamma/\kappa\rightarrow0$ as long as the conditions Eq. (14) is satisfied. This is important because, in general, mechanical decay rate $\gamma$ is much less than cavity decay rate $\kappa$ in cavity optomechanics. In addition, even with very weak optomechanical coupling $(G\ll\kappa)$, perfect optical nonreciprocity can still occur as $\gamma\ll\kappa$ according to Eq. (14).

In Fig. 2(a)--2(d), we plot transmission amplitudes $T_{LR}$ (red line) and $T_{RL}$ (black line) vs normalized detuning $x/\kappa$ with $J=\frac{\kappa}{2}$, $G=\frac{\sqrt{\kappa\gamma}}{2}$ for $\gamma/\kappa=2$, $1$, $1/5$, $1/100$ respectively. It can be clearly seen from Fig. 2 that mechanical decay rate $\gamma$ indeed does not affect the appearance of perfect optical nonreciprocity, but can strongly affect the width of transmission spectrum, especially for the case of $\gamma\ll\kappa$. The two curves of transmission amplitudes $T_{LR}$ and $T_{RL}$ tend to coincide outside the vicinity of resonance frequency ($x=0$) when $\gamma/\kappa\rightarrow0$. But the system always exhibits optical nonreciprocity near resonance frequency in the case, such as $\gamma/\kappa=1/100$ [see Fig. 2(d)]. By the way, the perfect optical nonreciprocity Eq. (9b) will occur if $\theta=\frac{\pi}{2}$.

\begin{figure}[b]
	\centering\includegraphics[width=0.42\textwidth]{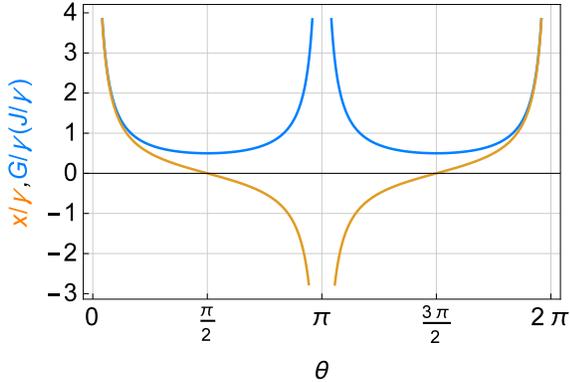}\caption{ Normalized coupling strengths $G/\gamma$ ($J=G$) (blue line) and detuning $x/\gamma$ (yellow line) are plotted vs phase difference $\theta$ according to Eq. (16).}%
	\label{Fig3}%
\end{figure}

\subsection{Equal damping rate $\kappa=\gamma$}

With equal damping rate $\kappa=\gamma$, the two optical output fields Eq. (10) now become
\begin{align}
\frac{\varepsilon_{R+}^{out}}{\varepsilon^{in}_{L}}&
=\frac{4\gamma(2e^{i\theta}G^{2}-iJ\gamma_{x})}{(8G^{2}+4J^{2}+\gamma_{x}%
	^{2})\gamma_{x}+16iG^{2}J\cos\theta},\nonumber\\
\frac{\varepsilon_{L+}^{out}}{\varepsilon^{in}_{R}}&
=\frac{4\gamma(2e^{-i\theta}G^{2}-iJ\gamma_{x})}{(8G^{2}+4J^{2}+\gamma_{x}%
	^{2})\gamma_{x}+16iG^{2}J\cos\theta}.\tag{15}%
\end{align}

\begin{figure}[b]
\centering\includegraphics[width=0.48\textwidth]{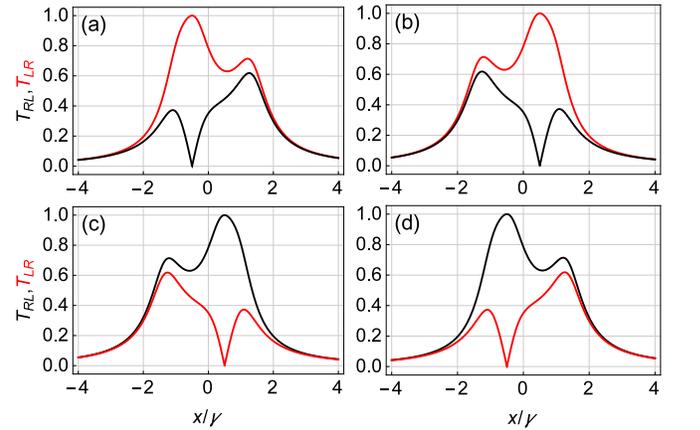} \caption{ Transmission amplitudes $T_{LR}$ (red line) and $T_{RL}$ (black line) are plotted vs normalized detuning $x/\gamma$ for different phase difference: (a) $\theta=-\frac{3\pi}{4}$, (b) $\theta=-\frac{\pi}{4}$, (c) $\theta=\frac{\pi}{4}$, and (d) $\theta=\frac{3\pi}{4}$. The coupling strengths $G=\pm\frac{\gamma\csc\theta}{2}$ ($J=G$) according to Eq. (16).}%
\label{Fig4}%
\end{figure}

From Eq. (15), we can obtain the conditions for perfect optical nonreciprocity as follows
\begin{align*}
x&=\pm\frac{\gamma\cot\theta}{2},\nonumber\\ J&=\pm\frac{\gamma\csc\theta}{2},\nonumber\\
G&=\pm\frac{\gamma\csc\theta}{2}\tag{16}
\end{align*}
where the negative sign and $\theta\in(\pi,2\pi)$ meet Eq. (9a), and positive sign and $\theta\in(0,\pi)$ meet Eq. (9b). It means that we can change the direction of isolation by adjusting the nonreciprocal phase difference $\theta\in(0,\pi)$ or $(\pi,2\pi)$.
In Fig. 3, we plot the normalized coupling strengths $G/\gamma$, $J/\gamma$ ($J=G$) (blue line) and detuning $x/\gamma$ (yellow line) vs phase difference $\theta$ according to Eq. (16). For the special case of $\theta=\pm\frac{\pi}{2}$, the coupling strength $G$ ($J$) takes the minimum value $\frac{\gamma}{2}$ and detuning $x=0$ (see Fig. 3), and the transmission spectrums $T_{LR}$ and $T_{RL}$ take a symmetric form with respect to detuning $x$ [see Fig. 2(b)].

From Eq. (16), we can see that perfect optical nonreciprocity can occur with any phase $\theta$ ($\theta\neq n\pi$) as long as $G\ll\omega_{m}$ ($|\sin\theta|\gg\frac{\gamma}{2\omega_{m}}$) where rotating wave approximation is valid. It means the strongest quantum interference takes place at detuning $x=\pm\frac{\gamma\cot\theta}{2}$ in the case of $\kappa=\gamma$.
In Fig. 4(a)--4(d), we plot the transmission amplitudes $T_{LR}$ (red line) and $T_{RL}$ (black line) vs normalized detuning $x/\gamma$ with  $G=\pm\frac{\gamma\csc\theta}{2}$ ($J=G$) for $\theta=-\frac{3\pi}{4}$, $-\frac{\pi}{4}$, $\frac{\pi}{4}$, $\frac{3\pi}{4}$, respectively.
It can be seen from Fig. 4, the transmission spectrums $T_{LR}$ and $T_{RL}$ will not take the symmetric form anymore as $\theta\neq\pm\frac{\pi}{2}$, and $T_{LR}>T_{RL}$ for $\theta\in(-\pi,0)$, $T_{LR}<T_{RL}$ for $\theta\in(0,\pi)$.

\section{conclusion}

In summary, we have theoretically studied how to achieve perfect optical nonreciprocity in a double-cavity
optomechanical system. In this paper, we focus on the conditions where the system can exhibit perfect optical nonreciprocity. From the expressions of condition, we can draw three important conclusions: (1) when nonreciprocal phase difference $\theta=\pm\frac{\pi}{2}$, the mechanical damping rate has no effect on the appearance of perfect optical nonreciprocity as long as Eq. (14) is satisfied; (2) even with very weak optomechanical coupling $(G\ll\kappa)$, perfect optical nonreciprocity can still occur according to Eq. (14); (3) the system can exhibit perfect optical nonreciprocity with any nonreciprocal phase difference $\theta$ $(\theta\neq0,\pi)$ if $\kappa=\gamma$ and Eq. (16) is satisfied. Our results can also be applied to other parametrically coupled three-mode bosonic systems, besides optomechanical systems.

\begin{acknowledgments}
L. Yang is supported by National Natural Science Foundation of China (Grant No. 11804066), the China Postdoctoral Science Foundation (Grant No. 2018M630337), and Fundamental Research Funds for the Central Universities (Grant No. 3072019CFM0405).
\end{acknowledgments}

\end{document}